# SD$^2$: Slicing and Dicing Scholarly Data for Interactive Evaluation of Academic Performance

Zhichun Guo, Jun Tao*, *Member, IEEE*, Siming Chen, *Member, IEEE*,
Nitesh V. Chawla, *Fellow, IEEE*, and Chaoli Wang, *Senior Member, IEEE*

**Abstract**—Comprehensively evaluating and comparing researchers' academic performance is complicated due to the intrinsic complexity of scholarly data. Different scholarly evaluation tasks often require the publication and citation data to be investigated in various manners. In this paper, we present an interactive visualization framework, SD$^2$, to enable flexible data partition and composition to support various analysis requirements within a single system. SD$^2$ features the hierarchical histogram, a novel visual representation for flexibly slicing and dicing the data, allowing different aspects of scholarly performance to be studied and compared. We also leverage the state-of-the-art set visualization technique to select individual researchers or combine multiple scholars for comprehensive visual comparison. We conduct multiple rounds of expert evaluation to study the effectiveness and usability of SD$^2$ and revise the design and system implementation accordingly. The effectiveness of SD$^2$ is demonstrated via multiple usage scenarios with each aiming to answer a specific, commonly raised question.

---◆---

## 1 INTRODUCTION

Accurately assessing the scholarly performance of a researcher plays a vital role in many important decision-making tasks, such as hiring a postdoctoral researcher or a tenure-track faculty member, evaluating a candidate for tenure promotion, and comparing a researcher's performance against his/her peers. Nowadays, many colleges and universities rely on publicly available online resources such as DBLP and Google Scholar (GS) when checking a candidate's scholarly credentials.

DBLP shows researchers' scholarly output majorly in textual format. Besides this, GS can also make basic statistical analyses, such as the individual's total citation count, $h$-index, and $i10$-index. These metrics can reflect the research outcome or impact in a specific aspect. However, the requirements of evaluating scholarly performance are much diverse and complicated, and these tools become limited as they may not facilitate the discovery of interactions among multiple factors. For example, when hiring a tenure-track faculty, other than the number of papers and citations, we may also consider the quality of his citations, such as whether his paper receives citations from the top venues. We may be interested in his independent academic performance as well. In this case, analyzing the papers that are not co-authored with his advisors may be desired. For another example, when searching for an academic advisor, a student

• Z. Guo was a visiting scholar at Sun Yat-sen University when the work is developed. She is with the Department of Computer Science and Engineering, University of Notre Dame, Notre Dame, IN, 46556. E-mail: zguo5@nd.edu.
• J. Tao is with the School of Computer Science and Engineering, Sun Yat-sen University and the National Supercomputer Center in Guangzhou, China. E-mail: taoj23@mail.sysu.edu.cn. He is the corresponding author.
• S. Chen is with the School of Data Science, Fudan University, China. E-mail: simingchen@fudan.edu.cn.
• N. V. Chawla, and C. Wang are with the Department of Computer Science and Engineering, University of Notre Dame. E-mail: {nchawla, chaoli.wang}@nd.edu.

may want to investigate the scholarly performance of the potential advisors' students, which may require groups of researchers to be compared. These questions can not be answered by the popular tools (e.g., DBLP and GS).

Different roles and analysis purposes may result in a variety of requirements in evaluating the scholarly performance, or various granularity to investigate the publication and citation data for details. Our research goal is to provide a visual analytic tool that provides sufficient flexibility to slice and dice the scholarly data to satisfy users' evaluation and comparison needs. The research challenges are as follows: **First**, how to design a tool that can interactively breakdown the information using the desired attributes to an appropriate level to answer various kinds of questions? **Second**, how to provide intuitive visualization for users to view and visually compare the information to glean insights? **Third**, how to automatically collect the rich set of publication and citation data that is required to support comprehensive evaluation?

To address these challenges, We present *SD$^2$*, a visual analytics framework for interactive evaluation of scholarly performance. The development of SD$^2$ stems from the familiar yet unaddressed needs of objectively and comprehensively evaluating the scholarly output and impact of researchers. To fully understand the evaluation requirements, we collect the opinions from recruitment committees of multiple departments at top research universities, graduate students, visual analytic experts, and data scientists. Our design and contributions are as follows:

*Our major contribution is the problem formulation*. We unify the broad spectrum of questions regarding the scholarly data into a problem of exploring multiple paper sets with attributes. With this unified formulation, the research outcome becomes the set of papers, and the research impact becomes the set of citations. The relations among researchers or even researcher groups can be flexibly modeled by using set oper-ations to combine the corresponding sets. The investigation of different factors can be performed by subdividing the sets based on the attributes of elements.

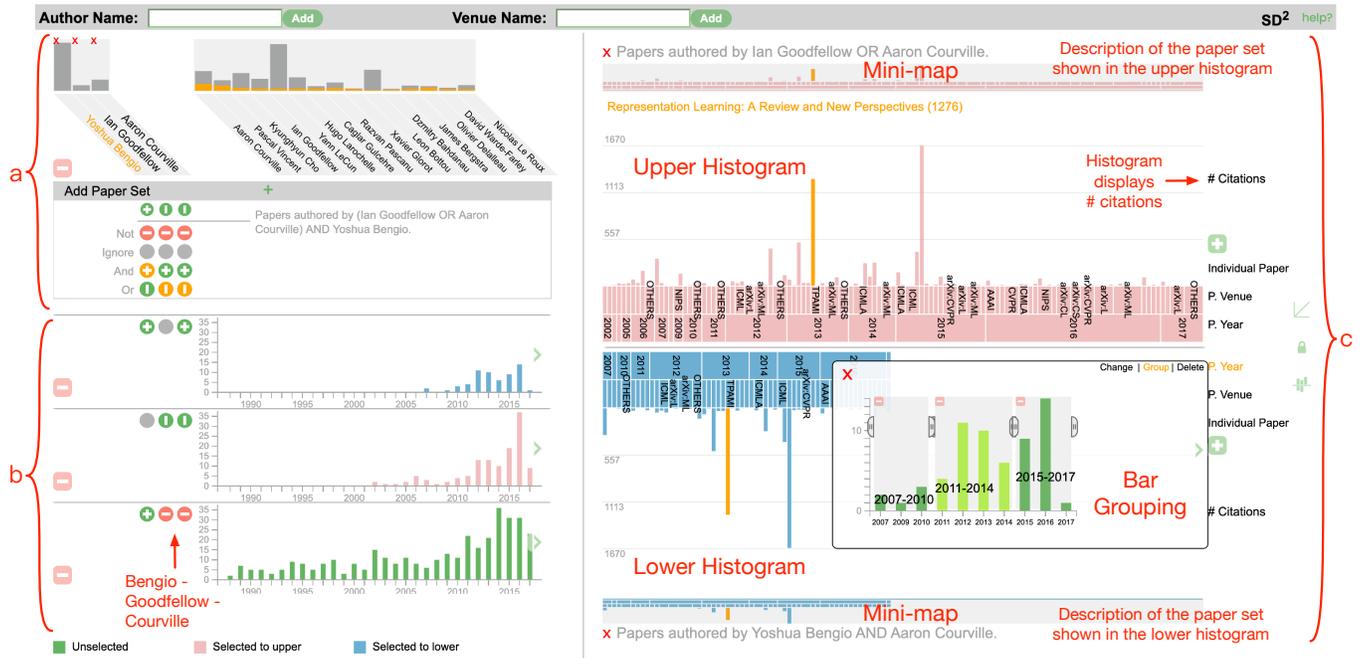

Fig. 1: SD$^2$ consists of three coordinated views: (a) scholar view, (b) publication view, and (c) hierarchical histogram view. The example shows a 2019 ACM Turing Award winner Yoshua Bengio's papers with his co-authors Ian Goodfellow or Aaron Courville. In (c), the upper histogram shows papers authored by Goodfellow or Courville and the lower histogram shows papers authored by Bengio and Courville.

*Our second contribution is the visual representation closely intertwined with the problem formulation.* Specifically, we design a new visual representation called the *hierarchical histogram* to support the flexibly slicing and dicing of scholarly data and enable easy observation and comparison of the hierarchies of data. This representation provides features useful in our scenario that are not available in similar designs (Section 6.3).

*We also contribute to the data preparation for egocentric exploration of scholarly data.* We combine Microsoft Academic Graph (MAG, a freely available bibliography database) for data acquisition and Google Scholar for resolving ambiguity. This minimizes the need for data cleaning, which is often costly in understanding scholarly data and allows the data to be crawled on the fly during the interaction. The source code is provided for public access.

## 2 RELATED WORK

In this section, we review related work in scholarly data visualization and general information visualization techniques used in our design. The scholarly data visualization techniques are studied as they target the same domain problems as our approach. The visual comparison of hierarchies and the egocentric exploration techniques are studied as our tool also follows similar design principals.

**Scholarly data visualization.** Scholarly data has been extensively studied in the past decades [20]. The existing research investigated different aspects of the scholarly data. Some research focused on the citation network. These approaches often design specific graphs or tree visual repre-sentations to show citation relationships. van Liere and de Leeuw [31] proposed GraphSplatting that converted the citation graph into a density field to highlight the central topics. Zhang et al. [38] visualized the paper-reference matrices using Frequent Pattern trees (FP-trees). FP-tree emphasizes co-citation relationships by placing the papers that are cited together close to each other. Shi et al. [25] proposed VEGAS that aggregates papers as nodes to summarize the citation network. Maguire et al. [22] visualized the paper based on its reference and the papers citing it. This visualization aims to demonstrate the impact of individual papers instead of the entire network. Shin et al. [26] presented impact flower, an egocentric visualization depicting the citations between an academic entity and others.

Some research aimed at discovering the evolution pattern of topics. These approaches feature multiple views for multi-faceted information regarding time, scholars, and topics. For example, Lee et al. [14] designed PaperLens, which shows the research trend of popular topics with bar charts and top-cited papers and authors as a grid of squares. Rind et al. [24] presented PubViz, which also shows the research trend using bar charts. It further highlights paper types, authors, and keywords in the corresponding year, when one of the bars is selected. Heimerl et al. [10] designed CiteRivers that visualizes paper clusters and the corresponding keywords as a streamgraph to study the research trend. Wang et al. [33] designed VISPubComPAS that shows topics of IEEE VIS papers of affiliations and authors. It visually compares two groups of scholars' research interest evolution through multiple bar charts, each corresponding to a topic. Keshif [37], although not specially designed for scholarly data, is similar to these techniques as it employs multiple views to present data from multiple dimensions. Dork et al. [5] presented PivotPaths that visualizes authors, papers, and keywords as nodes in three disjoint regions and allows their connections to be interactively explored.

Some research studied the collaboration and impact of scholars. These approaches often adopt an egocentric design. Wu et al. [35] presented PathWay to visualize the career paths of researchers. The numbers of papers, collaborations, and challengers over years are shown by line charts and glyphs. It also supports comparison among researchers. Wang et al. [32] presented ImpactVis that adopts a matrix-based design to visualize the paper and citation of a researcher. Wu et al. [36] designed egoSlider, an egocentric interface to study the evolution of collaboration networks of scholars. The networks of individual scholars can be summarized by several pie charts over time, and expanded to a graph on demand.

Our $SD^2$ shares similarity with the previous work in many aspects. For example, it reveals information along multiple axes, similar to PaperLens [14] and Keshif [37], and it adopts an egocentric design and supports comparison, similar to ImpactVis [32] and egoSlider [36]. However, our $SD^2$ also differs from the above work in multiple ways. First, the most notable difference is its diversity and flexibility. For example, when publication year is used to build the hierarchical histogram, it can be used to demonstrate the research trend; and, when venue rank is considered, it can facilitate the analysis of publication and citation quality. $SD^2$ also enables flexible combination of multiple researchers to study their collaboration patterns. Second, $SD^2$ can reveal the interactions among multiple factors. By changing the order of the factors in constructing the hierarchical histograms, users can switch the focus of factors and understand its relationships to other factors. ImpactVis [32] supports two factors (publication and citation years), but it does not allow the interactions among other factors to be discovered. Please refer to Section 9 for a detailed comparison of $SD^2$ with other techniques.

**Visual comparison of hierarchies.** We model the scholarly performance as papers and citations. Visual comparing the paper sets in a hierarchical manner is essential to comprehensively understand and compare the paper sets. While well-known information visualization techniques such as hyperbolic tree [13], treemap [27], and sunburst [28] enable the visualization of a large data hierarchy, they do not usually support a direct comparison of two different hierarchies. Further techniques are developed for visually comparing hierarchies. Examples include TreeJuxtaposer [23], contrast treemap [29], CandidTree [15], TreeVersity [9], nested icicle plot [2], PhenoBlocks [8], aggregated dendrograms [21], and BarcodeTree [18].

Most of the existing comparative techniques are built on top of the well-known hierarchy visualization, such as hyperbolic tree [13], treemap [27], and sunburst [28]. Our hierarchical histogram is similar to the icicle plot [12]. But instead of mapping the size of a node to the width of a block, we use a uniform width for all leaf nodes and represent the size of a leaf node using the height of a bar. In this way, both the numbers of children of internal nodes and the sizes of leaf nodes are visually encoded. Please refer to Section 6.3 for a detailed design choice discussion.

**Egocentric exploration techniques.** Instead of giving an overview, $SD^2$ focuses on personal visualization using an egocentric approach: users select researchers of interest and compare their scholarly performances. Examples of egocentric visualization include Episogram [3], egoSlider [36], EgoLines [39], egoComp [19], and LikeMeDonuts [6]. Fung et al. [7] presented a comparative study on personal visualizations of bibliographic data using three designs: node-link diagrams, adjacency matrices, and botanical trees. They found that node-link diagrams are good at revealing over distributions, adjacency matrices can convey more information with less clutter, and botanical trees provide the best at a glance characterization. To allow egocentric visualization and facilitate direct comparison, we prefer a timeline-based design over a node-link diagram or radial layout. Our simple design leverages the familiar form of bar charts to effectively convey aligned information.

## 3 TASK ANALYSIS

A motivating example is that when evaluating an applicant for a professorship or promoting a tenure-track professor, the searching committee always wants to evaluate the candidate from multiple perspectives.

The searching committee would first ask these questions: *is the researcher still actively publishing*, *where does the researcher publish*, and *what are the research fields or topics of interest of the researcher over time*? The quantity of papers indicates the productivity and activeness of a researcher, and the publication venues reflect the quality of research to a certain degree. Thus, the first task to evaluate a researcher's performance is to **investigate his/her publication records by attributes (T1)**.

The citation record of a researcher is a major indicator of his/her research impact. The total number of citations is often used to evaluate the overall impact of a researcher, but a single highly-cited paper could skew it. The $h$-index is more robust as it considers both the numbers of citations and papers. However, it still lacks the fine granularity of information for critical decision-making for hiring new faculty members. For example, the searching committee would want to reveal information related to *temporal influence patterns* (e.g., does the citation number increase or decrease over the years?), *domain-specific influence* (e.g., what are the major research fields that cite the researcher?), and the *quality of citation* (e.g., is the researcher cited by high-impact papers at top venues?). Indeed, the traditional metrics will fail to answer more complicated questions like *is the researcher getting more high-impact citations in a specific field*? This information can be crucial in identifying active and/or influential faculty candidates in a targeted research field. Thus, the second derived task is to **understand a researcher's citation record by attributes (T2)**.

Independence is essential for a mature researcher after his/her Ph.D. graduation. It indicates that the researcher can perform research independently without relying on others. Usually, the searching committee would compare one's publication and citation records with specific collaborators (e.g., one's advisor) against the records without them. For example, *how many papers are published by the researcher without his/her advisor*, and *how many citations come from these papers*? Additionally, breakdown information over time and research topics should also be provided to understand the record from desired aspects. For example, the committee wants to know *does the researcher rely on certain collaborators to publish on a specific topic*, and *do the researcher's papers have a similar impact without these collaborators*? Thus, the third task can be summarized as **evaluate the independence of a researcher (T3)**.

The searching committee usually has a list of candidates. Peer comparison provides contextual information to evaluate one's scholarly performance and predict one's potential impact. Comparing *two researchers in the same field at a similar career stage* provides the most direct answers to these questions: *does one researcher publish more papers than the other, is one more independent or influential than the other*, and *does one demonstrate a more promising trend in terms of papers and citations than the other*? When more references are needed, it is also common to compare *two researchers in related fields at a similar career stage*. So, the difference between the fields should be addressed by the tool. Additionally, to predict one's future impact, it will be beneficial to compare *a junior researcher against a senior one in the same field*. We want to know whether the junior one has comparable performance as that of the senior one during his/her early career, or whether the junior has a comparable trajectory as that of the senior. Thus, the fourth task is **compare individual researchers (T4)**.

When the candidates have their research groups, search committees may also be interested in the research outcome of their groups. Besides, when evaluating two senior researchers' advising records, the committee may compare their previously graduated students as groups. For example, they may ask questions: *whose students have more papers and citations during their graduate study, whose students have a better aggregate record after graduation*, and *whose students are more independent after graduation*? Thus, the task can be summarized as **compare two groups of researchers (T5)**.

Furthermore, these tasks may also apply in other scenarios. For example, when a prospective student is searching for an academic advisor, she may want to know *is this professor is still active* (**T1**), *do the students of this professor have a good publication and citation record* (**T1** and **T2**), and *do the students of this professor outperform the students of that professor* (**T5**)? Similar questions may be asked when electing fellows or searching for collaborators, and these tasks will help answer the questions.

## 4 DESIGN REQUIREMENTS

We aim to design a visual analytics tool to support the tasks mentioned above under two design principles: *lightweight* and *flexible*. On the one hand, "lightweight" indicates that the tool should be efficient with a minimum amount of computation involved. Instead of mining and analyzing the entire academic network, we target the local analysis of user-specified researchers, including investigating individual researchers, understanding their collaboration, and comparing multiple researchers. On the other hand, "lightweight" indicates that the tool should use simple and common visual representations requiring minimum learning effort from users. By "flexible", we mean that the tool should be customizable to answer all questions raised in

Section 3 in a unified way. Under these two principals, we elaborate on our design requirements as follows:

**R0. Hierarchical exploration with visual scent cues.** The tool should provide information on multiple levels of detail, allowing users to drill down to the desired publication (**T1**) and citation (**T2**) information gradually. At each level, additional information should be provided as visual cues to guide further navigation, as inspired by scented widgets [34]. The specific tasks to be performed at each level (coarsest, intermediate and finest) and the visual scents to be presented will be explained in further detail in the later requirements.

**R1. Specifying researchers of interest and generating overview.** At the coarsest level, users should be able to specify researchers for investigation, and the tool should generate an overview of their papers (**T1** and **T2**) and collaboration record (**T3**). The overview should provide the high-level information of a researcher, e.g., *who are the most frequent collaborators of the researcher*, and *how many papers does the researcher publish with each collaborator, respectively*? The overview will assist users in selecting additional researchers or specify how the information of existing researchers should be combined for further investigation.

**R2. Combining of record from multiple researchers and revealing relationship.** At the intermediate level, users should be able to combine the record of existing researchers to reveal the desired relationships (**T4**). For example, consider two researchers, $A$ and $B$. Users should be able to generate *the record with both A and B to examine their collaboration* (**T3**), *the record with A but without B to evaluate the independence of A* (**T3**), and *the record with either A or B to investigate their combined impact* (**T5**). Similar rules should be allowed to combine more researchers. At this level, the tool should provide temporal patterns of the combined record (e.g., the paper numbers over the years) for users to verify that the combination is indeed meaningful.

**R3. Partitioning the information in multiple ways and showing the details.** At the finest level, users should be able to slice and dice the information so that different aspects of the publication and citation record can be revealed and studied (**T1** and **T2**). For example, when a collection of papers is first partitioned by the topics and then by the publication years, we may find out how the research interest of the corresponding researchers changes over time (**T1**). When a collection of citations is first partitioned by the topics of citing papers and then by their citation years, we may find out how the corresponding researchers' influence on different topics evolves (**T2**). When a collection of citations is first partitioned by the venue ranks of reference and then by the ranks of citing papers, we may verify whether the papers published at better venues receive more citations from top venues (**T2**), etc. Leveraging the rich information associated with the papers, we should be able to answer questions from diverse aspects.

**R4. Aligning partitioned data for comparison.** The comparison between two collections of papers should be performed at the finest level using the partitioned information. Individual comparison is supported when each collection is produced from a single researcher (**T4**), and group comparison is enabled when each collection corresponds to a group of researchers (**T5**). The visual representation should allow the information to be aligned so that the corresponding parts can be easily compared. Both automatic and manual alignment should be provided. Automatic alignment should coordinate the same items in the two collections, while manual alignment should allow users to customize the alignment to match items carrying similar meanings. For example, to address the difference between different career stages, users should be able to dynamically align the time axes of two researchers.

## 5 DATA PROCESSING AND HANDLING

Obtaining a comprehensive data set of cleaned publication reference is challenging due to the inherently noisy citation data. To address this issue, manually labeling or correction might be involved in previous work [11], [32]. However, this may limit the size of data that can be processed. For example, `vispubdata.org` [11] opts to limit the citation relation to the IEEE VIS conference only.

Our data come from Microsoft Academic Graph (MAG) and Google Scholar (GS). MAG is a knowledge graph updated weekly. The version we use contains more than 164 million papers and one billion citation relationships. Each paper has its ID, title, keywords, abstract, authors and their institutions, venue name, topics, and fields of study. Each citation relationship is a pair of the reference paper ID and the citing paper ID. However, although MAG contains rich information, it provides limited power to distinguish researchers of the same name. In contrast, GS also maintains comprehensive citation records but suffers less from author name ambiguity as many authors retain their own paper lists. However, GS does not make its data open, which makes it impractical for large-scale analysis. In our experiment, we mostly rely on MAG for the publication and citation information. We convert all data in MAG into a SQL database for efficient retrieval. We only use GS to mitigate name ambiguity.

**Author name disambiguation.** We rely on GS to identify the list of papers authored by a specific researcher. Although GS is not entirely error-free, it is by far one of the most accurate data sources for this purpose. For each researcher to be examined, we crawl his/her paper list from GS. Then, we query the SQL database to obtain the information of each paper (including the papers citing this work and their information) from MAG. This avoids frequent queries to GS, which is prohibited by Google, but still connects the authors to their papers with reasonable accuracy.

**Venue name disambiguation.** The ambiguity among venue names mostly comes from editing errors and different styles of venue names. Levenshtein's edit distance [16] is often used to identify two names of the same venue [11]. This may be effective against small editing errors with a few characters, but it could be less capable of handling different styles. For example, "IEEE Transactions on Visualization and Computer Graphics", "Visualization and Computer Graphics, IEEE Transactions on", "IEEE Trans. Vis. Comput. Graph.", "IEEE TVCG", and "TVCG" all refer to the same journal, but the edit distance between them can be large. Since this paper does not target at resolving the ambiguity, we rely on Google to identify the correspondence among different names. We search the venue name in DBLP using Google and consider the names returning the same DBLP link to be the same venue. In the given TVCG journal example, all the five names point to the same link in DBLP.

**Venue classification.** For publication/citation venues, we follow the list of computer science conferences and journals recommended by China Computer Federation (CCF) [4]. The CCF recommendation list classifies 571 major venues into ten categories: system and architecture, networks, security, database and mining, software engineering, theory, computer graphics, artificial intelligence, human-computer interaction, and interdisciplinary. For each category, the venues are further divided into three ranks from A to C with rank A venues being the most prestigious ones. During exploration, we use the categories and ranks as additional attributes to group the venues.

**Implementation.** Our implementation consists of four major components: a MySQL database, a C++ connector, a Python server, and a web interface. The MySQL database stores the entire MAG data and responses to the C++ connector upon request. The C++ connector is implemented as a Python plugin. It connects the MySQL database and the Python server. It requests the MAG data from the MySQL database upon requests from the Python server. It caches the data under investigation, and it is responsible for processing the cached data (e.g., grouping based on attributes, counting the number of papers or citations, and computing $h$-indices, etc.). It avoids redundant access to the MySQL database if the requested data is already cached. The C++ connector deals with the computation-intensive requests to provide interactive performance. This performance is not available with a Python server when the amount of requested papers is large, which is often the case when we group the citation papers. The Python server is implemented based on Flask. It prepares data for the web interface. It crawls the GS website if the GS data is needed, and it calls the functions in the C++ connector if the MAG data is needed. The web interface is implemented based on D3. It interacts with users and sends requests to the Python server. Our source code is available at https://github.com/zhichunguo/SD2.

## 6 SD²: VISUALIZATION DESIGN

$SD^2$ follows the interaction model for large graph exploration: "*search, show context, expand on demand*" [30]. As shown in Figure 1, the interface of $SD^2$ consists of three tightly-coupled views, each of which corresponds to one step in the interaction model (**R0**). Namely, the three views are: (a) the *scholar view* ("*search*"), (b) the *publication view* ("*show context*"), and (c) the *hierarchical histogram view* ("*expand on demand*"). Typically, users will start by adding scholars of interest in the scholar view (**R1**), as shown in Figure 1 (a). They may further specify set operations to combine the papers of the existing scholars (**R2**) so that the desired relationships can be studied. Then, in the publication view, users can see how the number of papers changes over time for each paper set, as shown in Figure 1 (b). Finally, users can add the paper sets of interest to the hierarchical histogram view for detail exploration, as shown in Figure 1 (c). They can customize the hierarchical histogram by specifying attributes at each level to reveal the information of their interest (**R3**). Users can add two paper sets for detailed comparison as well (**R4**).

**Notations.** We define a paper to be a $k$-tuple of its attributes, i.e., $p = (a_1, \ldots, a_i \ldots, a_k)$, where $a_i$ is the $i$-th attribute of a paper, and $k$ is the number of attributes. In this paper, we consider the following attributes for a paper: title, venue, publication year, number of citation received, and $h$-index. The $h$-index of a paper is computed from all papers citing this one (i.e., the $h$-index of the collection of citing papers), serving as an alternative indicator of impact other than the citation count. For simplicity, we may refer to an attribute $a_i$ of a paper $p$ as $p(i)$.

A citation relationship is a link between two papers. Similarly, we represent a citation link using a $2k$-tuple, i.e., the $k$ attributes of the citing paper together with the $k$ attributes of the cited one. This allows citation relationships and papers to be handled in the same manner.

We denote a scholar as the set of papers authored by him/her, i.e., $\mathbf{S} = \{p_1, \ldots, p_m\}$. In this way, the papers of a scholar $\mathbf{S}_i$ without another scholar $\mathbf{S}_j$ is denoted as $\mathbf{S}_i - \mathbf{S}_j$, the papers co-authored by $\mathbf{S}_i$ and $\mathbf{S}_j$ is denoted as $\mathbf{S}_i \cap \mathbf{S}_j$, and the papers authored by either $\mathbf{S}_i$ or $\mathbf{S}_j$ is denoted as $\mathbf{S}_i \cup \mathbf{S}_j$.

Next, we describe the scholar view and the hierarchical histogram view in detail. We omit the discussion of the publication view, as its visual representation and interaction are relatively straightforward.

## 6.1 Scholar View

The scholar view is used to select scholars as paper sets and specify the set operations to "combine" these paper sets. The design of this view is inspired by UpSet [17]. The upper panel of this view shows the scholars under exploration (left) and the co-authors of the scholar of focus (right), as shown in Figure 1 (a). The ordered list of co-authors is obtained from the author's Google Scholar profile. The bar charts show the number of papers authored by each scholar. For each of the co-authors, we further display an orange bar to indicate the number of co-authored papers with the scholar of focus. For the example given in Figure 1 (a), the orange bars indicate that more than half of Pascal Vincent's papers are co-authored with Yoshua Bengio, and Yoshua Bengio also collaborates with Aaron Courville frequently. Users can click on any scholar under exploration to switch the focus, so that his/her co-authors will be displayed on the right. Users can click on any co-author to add them to the selected list as well.

To add a paper set for further examination, users can combine the scholars using set operations provided in the lower panel, as shown in Figure 1 (a). This panel supports four operators for each scholar: namely, "*not*", "*ignore*" "*and*", and "*or*". The design is similar to UpSet but with an additional "*or*" operator to allow the union of scholars, which is essential for studying the combined research output. The default operator for a scholar is "*ignore*", meaning that the scholar is irrelevant to the paper set being generated. Users can specify any other operator to get a scholar involved. A textual description of the resulting paper set will be updated whenever an operator is changed.

The paper set produced by this panel is the papers authored by any scholars labeled by "*or*" intersected by the papers co-authored by all scholars labeled by "*and*" excluding the papers authored by any scholars labeled by "*not*".

For the example given in Figure 1, from top to bottom, the three histograms shown in the publication view represent "Bengio ∩ Courville" (which can also be denoted as "Bengio + Courville"), "Goodfellow ∪ Courville" (which can also be denoted as "Goodfellow | Courville"), and "Bengio − Goodfellow − Courville", respectively. Note that our format does not support all possible forms of equations using set operations, but it can facilitate answering any questions raised in Section 3.

## 6.2 Hierarchical Histogram View

The hierarchical histogram view allows users to break down a set of papers or citations into a hierarchy of bars to answer specific questions. Without loss of generality, we describe our approach using paper sets. The interface of the hierarchical histogram view is shown in Figure 1 (c). Up to two paper sets can be added in this view as diverging bar charts for efficient comparison: one will be mapped to the *upper* histogram and the other to the *lower* histogram. Unlike the traditional bar charts, which list all bars of a histogram at a single level, the hierarchical histogram features a multilevel design: *several* levels of *horizontal* bars at the *intermediate* levels of the hierarchy where the *width* of each bar indicates the corresponding number of leaf nodes, followed by *one* level of *vertical* bars at the *finest* (i.e., leaf) level of the hierarchy where the *height* of each bar can indicate the number of papers, the total number of citations received by the papers, or the $h$-index of the papers.

To construct the hierarchy of a paper set, users can specify a series of attributes to partition the original set. At each level, an attribute will be applied to further partition the existing bars at this level into multiple smaller ones. Formally, consider a bar containing $n$ papers $\mathbf{P} = \{p_1, \ldots, p_n\}$ and an attribute $a_j$ for partitioning. The partitioning process will create multiple smaller paper sets whose union equals the original set, i.e., $\mathbf{P} = \cup \mathbf{P}_i$, where each partitioned set $\mathbf{P}_i$ consists of papers with the same value $v$ of $a_j$, i.e., $\mathbf{P}_i = \{p_k | p_k(j) = v\}$. In our implementation, we restrict the maximum partition level to four, as the interactions among more than four attributes will be difficult to understand.

We provide a series of publication attributes and an extra set of attributes for their citations. The publication attributes include citation counts, publication years, venues, CCF ranks, and titles (for distinguishing individual papers), and citation attributes are provided similarly. On the interface, the publication attributes and citation attributes are denoted using the "P." and "C." prefixes, respectively. An attribute can be changed or deleted when clicked on. Users can also swap the partition order of the paper set by dragging an attribute up or down.

For example, in Figure 2, the publication CCF rank ("P. CCF Rank", first level) and the publication year ("P. Year", second level) are used to partition Huamin Qu's publication data. Figure 2 (a) shows his publication trend in venues of different ranks by visualizing the number of papers in each bar. A clear increasing trend of his papers in the CCF rank A and non-ranked venues can be observed. Note that many venues are not ranked because they are relatively young, or they are not in the computer science field. Figure 2 (b) reveals the temporal citation pattern by showing the number of citations. The number of citations is visualized using the same partitioning attributes. Clearly, the papers published in the CCF rank A venues receive more citations, and the papers published from 2007 to 2011 are highly cited.

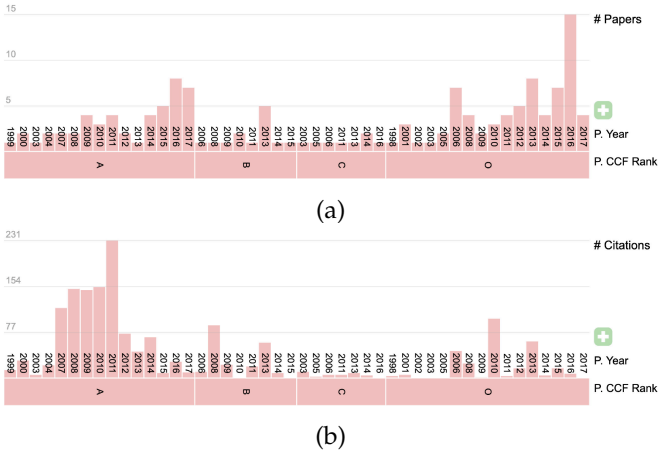

Fig. 2: Hierarchical histograms visualizing Huamin Qu's (a) publication and (b) citation data organized by their publication years and CCF Ranks.

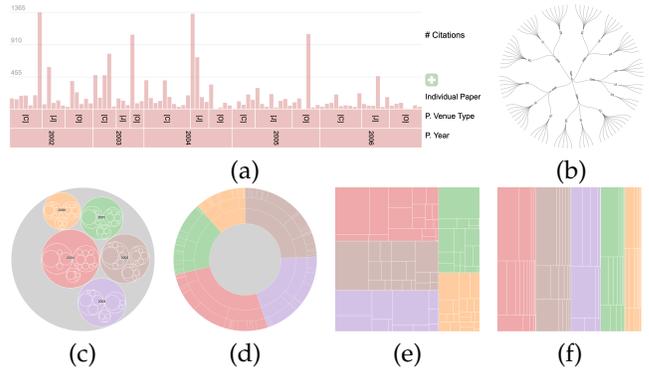

Fig. 4: Alternative designs for visualizing hierarchical data. (a) to (e) show our hierarchical histogram, radial tree, circle packing plot, sunburst diagram, squarified treemap, and slice-and-dice treemap, respectively.

**Bar grouping.** We provide the *bar grouping* feature, which allows users to manually group the bars to further reduce visual complexity. Figure 1 (c) illustrates such an example. Users can brush the histogram of the "P. Year" attribute to create periods of years. In the resulting hierarchical histogram, each period, instead of each year, forms a bar. Once a bar group is formed, users can also remove it (acting as a filter). For example, users can update the "P. CCF Rank" attribute to form bar groups and remove the one containing papers that are not in CCF rank A, leaving only CCF rank A papers to be shown and explored in the hierarchical histogram. To remove a group, users can either rename the group as "ignore" or simply click the button with a minus sign at the upper left corner of the group.

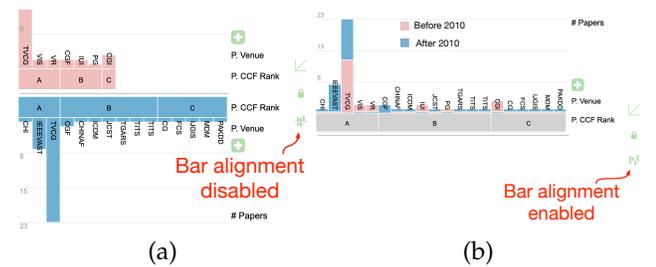

Fig. 3: Visual comparison of Huamin Qu's paper numbers at different venues before and after 2010. (a) shows the histograms without bar alignment and (b) shows the histograms with bar alignment. The aligned upper histogram shows the publication data before and after 2010.

**Attribute lock and bar alignment.** To facilitate efficient comparison of the upper and lower histograms, we further provide two features to coordinate the two histograms: *attribute lock* and *bar alignment*. The attribute lock feature aims to synchronize the attributes used to partition the data for the two histograms, so that the change of attributes in one histogram will be applied to the other automatically. The bar alignment feature aims to align the horizontal bars at the coarsest level in the two histograms. By default, each histogram shows a compact view of bars, where there is no space left between the bars. When bar alignment is enabled, the bars corresponding to the same attributes will be vertically aligned by inserting empty bars. In addition, all the bars will be displayed on the upper side, and the shorter bars will be overlaid on top of the corresponding taller bars. This design allow easier comparison of the quantity differences. Please refer to Section 8.5 for detailed design choice.

For example, Figure 3 compares Huamin Qu's paper numbers at different venues before 2010 and after 2010. In Figure 3 (a), without bar alignment, we can easily compare the number of publication venues, but it is difficult to identify where the differences are. In Figure 3 (b), with bar alignment, we can now easily compare the quantity difference in each venue.

**Description of the paper set.** Textual hints are provide to remind users the content displayed in the upper histogram and lower histogram. The descriptions of the two paper sets are placed in the uppermost and lowermost parts of the window, respectively, as shown in Figure 1 (c). When bar alignment is enabled, the two description will be combined into one and shown on the top of upper histogram. The combined description is composed of three parts: the description of the upper paper set in red, the description of the lower paper set in blue, and a term "VS" in gray that connects the two descriptions.

**Mini-map.** We provide users with a mini-map for easier navigation of the bars. This feature will be useful when a large amount of bars are displayed at the top layer leading to small width of each bar. The mini-map is equipped with two operations: horizontal scaling and scrolling. The horizontal scaling allows users to adjust the width of bars, and the scrolling allows users to specify a range of bars for detailed exploration.

When both the attribute lock and alignment are disabled, the two mini-maps can be used independently to scale and scroll the corresponding histograms. When the attribute lock is enabled, the bars in both histograms carry similar meaning. Therefore, we link the scaling operation of both histograms to enforce the same width for all the bars at the top level. When the alignment is enabled, we further link the scrolling of the two histograms, meaning that scrolling one histogram will move the other histogram simultaneously.

### 6.3 Alternative Design Choices

Our interface should support two core features: *using set operations to combine the research outcome of multiple researchers* (**R2**) and *visually exploring and comparing the paper sets with hierarchies* (**R3** and **R4**).

**Why UpSet [17]**? For the set operations (**R2**), we opt to use UpSet [17] for SD$^2$ as it is scalable in terms of the number of sets. Users only need to select a single operation for each scholar to specify how his/her paper set would be combined with others'. This design reduces visual complexity and interaction effort, while still maintaining the ability to produce the combinations to fulfill the requirements stated in Section 4.

**Why hierarchical histogram**? For visualization and comparison of hierarchical paper sets (**R3** and **R4**), we compare our hierarchical histogram against other alternatives using a paper set with three grouping levels in Figure 4. First, hierarchical histograms use a 1D subdivision for a straight layout similar to icicle plots [12], which better preserve the order of elements and groups (e.g., P. Year in Figure 4). Sunburst diagram [28] and radial tree also preserve the order along a circle, but the starting and ending years are more difficult to be identified. Slice-and-dice treemap preserves the order in alternating dimensions. Therefore, users have to switch from row and column pictures to identify the groups at different levels. Circle packing and squarified treemap do not preserve the order. Squarified treemap seems to be less effective in visualizing multiple levels of hierarchies. For example, it is difficult to tell how many subgroups exist in the brown group in Figure 4 (e). In addition, the straight layout of hierarchical histograms allows effective alignment of elements and groups for upper-lower comparison. Circle packing and sunburst diagram use circular layouts, where different layers vary in arc-lengths. This renders it less compelling to map two paper sets on different layers for comparison. Slice-and-dice treemap switches the order alternatively, making it challenging to align groups with varying numbers of elements or subgroups. Circle packing and squarified treemap do not list the elements and groups in a certain order, which is even more difficult to be aligned and compared.

Second, hierarchical histograms visualize the number of elements in each group or subgroup (using its width) together with the sizes of individual elements (using the height of the bars). The alternatives do not encode both of these properties explicitly. Radial tree only shows the number of elements using the arc-lengths. In contrast, circle packing, sunburst diagram, and treemaps only show the elements' sizes using the circles' sizes, arc-lengths, and rectangles' sizes, respectively. The bar plot design of hierarchical histograms allows more efficient comparison of element sizes. For example, in Figure 4 (a), it is easier to identify the paper with the most citations or the papers with over 900 citations using hierarchical histograms than using other alternatives.

**Why upper-lower layout**? For comparison, we choose a upper-lower layout instead of using clustered bars (as Keshif [37] and VisPubComPAS [33]) and stacked bars for the following reasons. First, clustered bars and stacked bars may need to use the color channel for visual encoding, which may increase visual complexity. We currently use red and blue colors to build connections between the publication view and the hierarchical histograms and use orange color to highlight mouse-hover bars. Using colors to encode additional information may lead to inconsistency. Second, clustered bars are similar to creating another attribute for different paper sets in our current hierarchical histograms. This may further increase the hierarchical histogram's width, leading to a high aspect ratio and inefficient use of screen space, especially when multiple attributes exist. Third, stacked bars are similar to our design, which divides the bars vertically. It may be useful to extend our comparison to multiple (groups of) scholars without the high aspect ratio problem. However, stacked bars may be less effective for a detailed comparison, as it lacks a unified baseline to align different categories. In contrast, our hierarchical histogram aligns two bars in the middle and use the reflection to show the difference.

## 7 USAGE SCENARIOS

In this section, we demonstrate the effectiveness and usability of SD$^2$ using several usage scenarios. We identify scholars in visualization, human-computer interaction (HCI), and data mining fields based on the Most Influential Scholar Annual List provided by AMiner [1]. In 2018, the winners were among the most-cited scholars whose papers were published in the top venues of their respective subject fields between 2007 and 2017. Recipients are automatically determined by a computer algorithm deployed in AMiner that tracks and ranks scholars based on citation counts collected by top-venue papers. As there are many interactions involved in the usage, we strongly encourage readers to watch the supplemental video for a more comprehensive understanding.

### 7.1 Scenario 1: Investigating Individual Researchers

This usage scenario mainly covers the following tasks **T1**, **T2**, **T3**, and **T4** stated in Section 3. We provide two ministudies with each aiming to answer a specific question.

#### 7.1.1 Who should I recruit or collaborate with?

When searching for a faculty candidate, it is not always the best strategy to recruit the one with more papers or citations. It is also important to study their independence, collaboration patterns, citation patterns, and their research interests from their prior record. Here, we examine two professors at their early career stage: Yingcai Wu and Nan Cao. Both of them graduated under the supervision of Huamin Qu and became professors at top research universities in China afterward. We first study the overview of their publication records with and without their advisor Qu (**T1**), as shown in Figure 5 (a). We can see that both of them have similar numbers of collaborated papers with Qu in recent years, but Cao has more independent papers (54) published than Wu (22) in general. However, by considering the citation records, we find that their performances are close: Wu's independent work has slightly more citations (257 against 241) and a slightly lower $h$-index (7 against 9). Therefore, to accurately evaluate their outcome without Qu, we need to investigate their profile details.

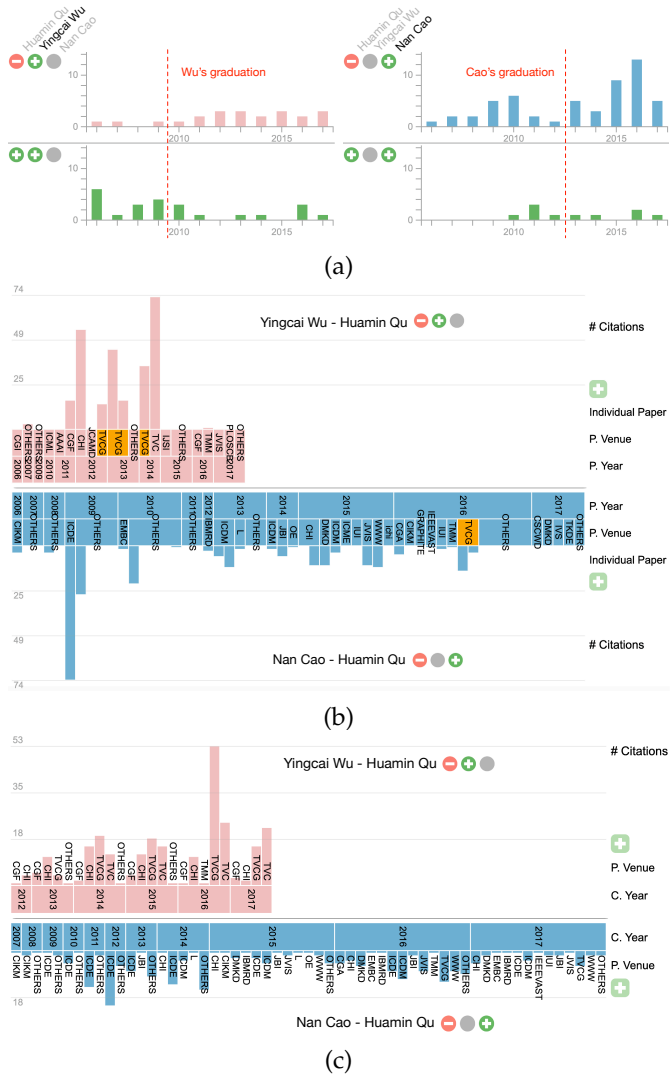

Fig. 5: Yingcan Wu and Nan Cao's independent and joint work with their PhD advisor Huamin Qu. (a) shows the numbers of papers of Wu over years in the first column, and those of Cao in the second. The first row shows the publication records without Qu and the second row shows those with Qu. (b) shows the numbers of citations of individual papers grouped by publication years and venues. (c) shows the numbers of citations of publication venues over citation years.

In Figure 5 (b), we use the hierarchical histogram to compare the publication and citation records of Wu without Qu (the upper histogram) and those of Cao without Qu (the lower histogram) (**T3** and **T4**). The heights of the bars represent the numbers of citations received by individual papers. The papers are further grouped by the citation venues and publication years. We find that both Wu and Cao published at top venues. But the papers of Wu focused more on the visualization or HCI. He had four papers that appeared in TVCG (as highlighted in orange), two in CGF, and one in CHI, TVC (The Visual Computer), and JVIS (Journal of Visualization). In contrast, Cao published in a broader range of venues. Other than the visualization and HCI, he also published in data mining and knowledge discovery venues, such as CIKM, ICDE, SDM, and WWW. In terms of citations, although the total numbers of citations for Wu and Cao are similar, the citation distributions are quite different. Wu had three papers with high citations and four with moderate citations, while Cao had one paper with high citations and eight with medium citations. Note that these are also the papers that contribute to the $h$-indexes of their independent work. In addition, Cao seems to receive more citations for his data mining papers.

We further study the temporal distribution of the citations over the publication venues (**T2**), as shown in Figure 5 (c). We can see that Wu's citations were mostly obtained for his papers in TVCG, CHI, and TVC, which did not change much over the years. But the number of citations increases steadily, especially for his TVCG papers. In contrast, Cao's citations concentrated on his work in ICDE from 2009 to 2014, and then gradually spread out to various venues in later years. These findings show that scholars may work in different styles and excel in diverse ways. According to the data we collected, Wu made steady progress in one direction and gradually built up his reputation in one field, while Cao adopted a different strategy. He tended to publish more papers in various fields, which may potentially increase his visibility and impact in the long term. For faculty hiring, the recruiters may need to evaluate the profiles of candidates comprehensively and make hiring decisions according to their ultimate expectations for the candidates.

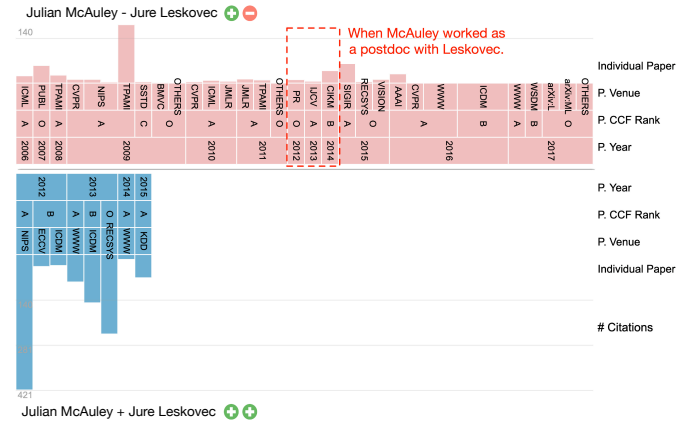

Fig. 6: This figure shows the detailed information of the papers published by Julian McAuley with (bottom) and without (top) Jure Leskovec, grouping by "Individual Paper", "P. Venue", "P. CCF Rank", and "P. Year". The red dashed-line rectangle highlights the period when McAuley was a postdoctoral researcher of Leskovec.

### 7.1.2 Will this advisor boost my career?

When seeking for a postdoctoral researcher position, one may wonder whether a potential advisor will boost his/her career (**T3**). To answer this question, it is often important to study the record of the advisor's former postdoctoral researchers, whenever available. Here, we investigate the record of Julian McAuley, who worked as a postdoctoral researcher under Jure Leskovec's supervision from late 2011 to 2014. We explore the detail of the two paper sets "McAuley − Leskovec" and "McAuley + Leskovec", as shown in Figure 6. The hierarchical histogram reveals the citations of individual papers successively grouped by publication years, venue ranks, and venue names. The upper histogram ("McAuley − Leskovec") shows that, without Leskovec's collaboration, McAuley still had a decent number of papers appearing at the top venues every year, such as NIPS, TPAMI, CVPR, and WWW, etc. His papers without Leskovec is roughly at the same level as their joint papers. What makes a difference is the number of citations. We find that their co-authored papers generally received more citations, as indicated by the taller bars in the lower histogram. Even within the paper set "McAuley − Leskovec" (upper histogram), we can see that three of the five most highly-cited papers were published after McAuley started his postdoctoral researcher position. Considering the significantly less time to accumulate citations, we speculate that McAuley's postdoctoral researcher experience grants him more chances to be cited, either due to the better visibility or more influential topics he is working on.

## 7.2 Scenario 2: Investigating Groups of Researchers

This usage scenario mainly covers the following tasks **T1**, **T2**, **T3**, and **T5** stated in Section 3. The success of senior professors does not only rely on the scholarly outcome of themselves but also on the success of their students. In this case, we would need to investigate the combined performance of several scholars as a group, and also factor in their collaboration with their advisors. Here, we use the two most influential scholars in the data mining field: Jiawei Han and Christos Faloutsos. Both of them have been working in the academy for more than twenty years and have graduated tens of PhD students. For concise analysis, we pick three representative alumni from each group for the study. The representative alumni are selected following three criteria: first, they are active in the academy; second, they are top co-authors of their advisors; third, they graduated in a similar range of time. We finally select Xifeng Yan (graduated in 2006), Deng Cai (2009), and Yizhou Sun (2012) from Han's previous students, Jimeng Sun (2007), Jure Leskovec (2008), and Hanghang Tong (2009) from Faloutsos' previous students. All these six scholars are listed among the top six co-authors in their respective advisors' Google Scholar(GS) profiles (accessed March 2019), and all of them currently hold tenure-track or tenured positions at top universities, fulfilling our selection criteria. For simplicity, in this section, we refer the papers authored by Han's students without Han as ("Han−") and with Han as ("Han+"), and the papers authored by Faloutsos's students without him as ("Faloutsos−") and with him as ("Faloutsos+").

**General observations.** We first compare their publication records (**T1** and **T5**). Figure 7 (a) shows that the overall patterns of paper numbers of "Han−" and "Faloutsos−" over the years are similar. This may due to the similar graduation time of the scholars being studied. For both groups of researchers, the numbers of papers started increasing to another level around 2010, which is approximately two years after their average graduation time. By grouping all the bars in the hierarchical histogram, we find that the total numbers of papers are close (346 for "Han−", and 387 for "Faloutsos−"). Figure 7 (b) further decomposes the papers into several groups according to their citation counts and CCF ranks for detail investigation, as highlighted by the red dashed-line rectangle. The histograms show that "Han−" had slightly more papers published in rank A venues, while "Faloutsos−" had more highly-cited papers with the numbers of citations larger than 50. But the overall papers' qualities are somewhat similar.

We then compare their citations (**T2** and **T5**). We find that the number of total citations of "Faloutsos−" (17897) is higher than that of "Han−" (7677), and the $h$-index indicates the same trend. By further grouping the citing papers according to their citation counts and CCF ranks, as shown in the blue dashed-line rectangle in Figure 7 (b), we find that "Faloutsos−" have more citations from the highly-cited papers and rank A papers as well, due to the larger number of total citations.

**Publication patterns over venues.** Finally, we compare their publication tendencies (**T1** and **T5**). Figure 7 (c) shows that the two groups of researchers share a large portion of their publication venues in common. This is not surprising as their primary research areas are similar. For example, both groups published most papers in two leading venues KDD and WWW. But we also notice that "Faloutsos−" was more focused on a relatively smaller number of venues, such as KDD, WWW, ICDM, and WSDM, while "Han−" had a more balanced distribution over the venues. Other than the appealingness of topics, publishing extensively in the same venue may be one potential factor to increase the citations as well, as this may help researchers gain extra visibility and possibly lead to more citations. But currently, we do not have more results to confirm this assumption as our tool is designed for supporting localized exploration.

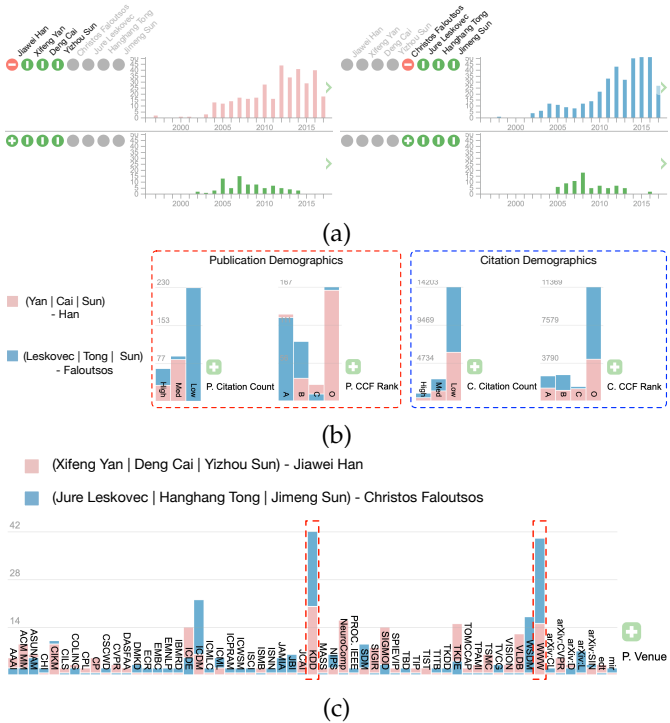

Fig. 7: Comparing two groups of researchers. (a) shows the numbers of papers for Jiawei Han's group in the first column and Christos Faloutsos' group in the second column. The first row shows the students' record independent of their respective advisors, and the second row shows the students' joint work with their respective advisors. (b) shows the numbers of papers grouped by the citation counts and CCF ranks in the red dashed-line rectangle, and the numbers of citations grouped by the citation counts and CCF ranks in the blue dashed-line rectangle. (c) shows the numbers of papers published in different venues, where two top venues (KDD and WWW) are highlighted in the red dashed-line rectangles.

In short, these usage scenarios demonstrate the capability of our approach for solving the practical questions through supporting the exploratory tasks of T1 - T5. In the next section, we demonstrate the usability and effectiveness of our proposed method by the evaluation from domain experts.

## 8 EVALUATION

The evaluation of SD² is performed in multiple rounds in various phase. In the development phase, two rounds of pivotal expert evaluations and a user study are conducted. In the revision phase, an empirical expert evaluation is used to examine the effectiveness of SD² with case studies and an additional round of user study is conducted to justify specific design choices. In general, the expert evaluations are used for in-depth interview regarding the effectiveness and overarching design principals, while the user studies examines usability of specific visual components and justify the detailed design choices.

The pivotal expert evaluation was performed with two visual analytics experts (postdoctoral researchers), one senior applied scientist at Microsoft research with years of experience in analyzing MAG, and a senior faculty running a bioinformatics research group of more than twenty graduate students and postdocs. These four experts evaluated SD² in various aspects, including visualization and interaction design, data analysis, and end-users' perspective. The user study recruited graduate students in computer science on their satisfaction with using SD² to complete the required tasks. Several revisions were made to address issues identified in the pivot expert evaluation, user study, and comments from the anonymous reviewers. Please refer to Section 8.4 in the main text for the user study, and Sections 1, and 2 in the Appendix for detail of the pivot expert evaluation and a complete history of tool revision, respectively.

After the revisions, we performed the second expert evaluation with two senior faculty members. Although researchers at different career stages may be potential users, we chose to evaluate the tool with senior faculty members for their broader view to analyze the data comprehensively. This evaluation was performed with three sessions. The evaluation started with a tutorial and a guided exploration using the usage scenarios in Section 7. Each of these two sessions took around 30 minutes. Then, the experts performed free-form explorations, which took each of them slightly more than an hour. We recorded their operations and conducted interviews with them. In this section, we present their operations and findings in the exploration and discuss the lessons learned. We omit the names of scholars studied at the experts' requests.

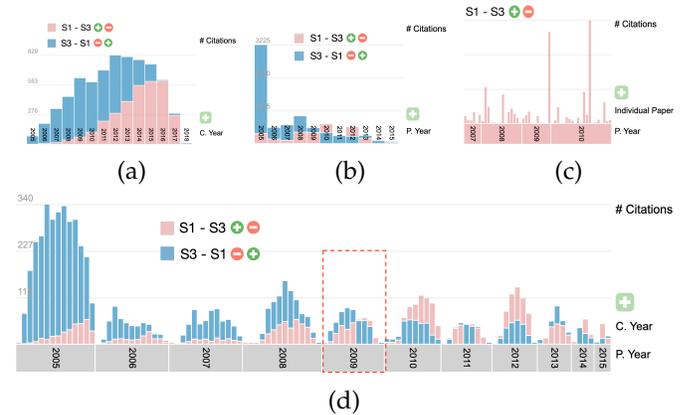

Fig. 8: Career path analysis results from Expert $E1$ to compare the independent work of two scholars $S1$ and $S3$. (a) compares the numbers of citations received by $S1 - S3$ and $S3 - S1$ over the citation years. (b) compares the numbers of citations received by $S1 - S3$ and $S3 - S1$ over the publication years. (c) shows the numbers of citations received by each paper of $S1 - S3$ over the publication years. (d) compares the numbers of citations received by $S1 - S3$ and $S3 - S1$ in each year over their publication years.

### 8.1 Career Path Analysis from Expert $E1$

Expert $E1$ is an established researcher in data science, whose research has received over 20 million dollars funding and more than 30,000 citations. $E1$ selected a scholar $S1$ at his mid-career stage for investigation. $E1$ first studied the performance of $S1$ with respect to $S1$'s advisor $S2$ (**T3**). He formed two paper sets $S1 + S2$ and $S1 - S2$ and compared them using hierarchical histograms. He found only one paper received high citations in $S1 + S2$, then he removed $S1 + S2$ from the hierarchical histogram and added $S2 - S1$ for comparison with $S1 - S2$. The relative performance of a scholar with respect to his advisor reduces the difference among research fields and topics. He used "P. Year" and "Individual Paper" as grouping criteria, but found that the hierarchies were difficult to compare as $S2$ had a much longer career. He tried to select papers over a ten-year period from 2005 to 2015 for comparison, as the MAG data used in our implementation contained papers up to 2017. He did not consider papers in 2016 and 2017 as these papers did not have enough time to demonstrate their impact. He found that selecting this period in the publication view would require an extra step to add them to the hierarchical histogram. Therefore, he opted to ignore the other years using the grouping function of the hierarchical histogram. He found the resulting visualization demonstrated a much clearer pattern that $S1$ published more high citation papers than $S2$. To study the general citation trend over recent years, he removed "Individual Paper" and found that the number of citations in each year received by $S1$ was higher than $S2$ as well. He commented that this indicated that $S1$ had become a successful independent researcher.

$E1$ then studied the performance of $S1$ with respect to $S1$'s department chair $S3$ (**T4**), as this reduced the differ-

ence in the working environment. He formed two paper sets $S1 - S3$ and $S3 - S1$ over the period from 2005 to 2015. This time, he specified the years of selection in the publication view before adding the paper sets to the hierarchical histogram. He used "C. Year" to see the citations received by the individual work of each scholar over the years. He found that although $S3 - S1$ received many more citations over the earlier years, the numbers of citations were similar after 2015, as shown in Figure 8 (a). He then used "P. Year" to see the citations received by each year's papers. He found that $S3$'s papers published in earlier years had more citations, but starting from 2010, $S1$'s papers outperformed $S3$'s (except for 2013) in terms of citation numbers, as shown in Figure 8 (b). He commented that this trend could not be found in Google Scholar(GS), as we could only view the numbers of citations received each year. He added an extra attribute "C. Year" to see the interaction of these two factors, as shown in Figure 8 (d). He found that the turning point was indeed 2009 instead of 2010, as highlighted in the red dashed-line rectangle. Although $S1$'s papers in 2009 received fewer citations than $S3$'s over the entire period, they started getting more citations from 2012. He commented that SD$^2$ was powerful to reveal the career path of rising stars in a department. However, during the evaluation, $E1$ did not study why 2009 became the turning point. By showing the citation numbers of individual papers, our post-analysis showed that $S1$ published his first highly-cited paper this year, as shown in Figure 8 (c).

$E1$ was interested in $S1$'s advising ability and studied the performance of $S1$'s students $S4$ and $S5$. He studied $S1 + S4$ and $S1 + S5$ and found that they had similar performance in terms of citation counts over the years. He stated that as $S5$ had several years of research experience before working with $S1$ while $S4$ did not, SD$^2$ was helpful to constrain the comparison scope to their collaboration with $S1$. $E1$ would like to explore their career development after graduation. However, as both $S4$ and $S5$ graduated after 2017, their publication data after graduation was not included in our data set. This is a limitation of our current tool.

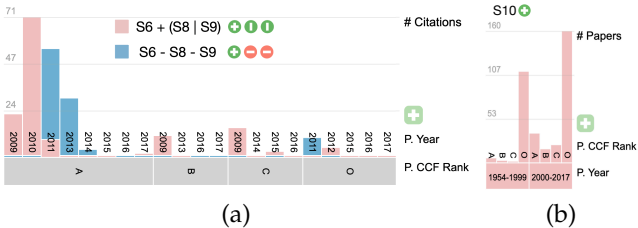

Fig. 9: Scholarly data studied by Expert $E2$. (a) shows the citations received by $S6$ with his advisors ($S6+(S8|S9)$) and without his advisors ($S6 - S8 - S9$) with two attributes "P. CCF Rank" and "P. Year". (b) shows the number of papers of $S10$ by their "P. CCF Rank" during two periods.

## 8.2 Publication Tendency Analysis from Expert $E2$

Expert $E2$ is a professor and the chief scientist at a super-computer center, leading a research group of more than twenty graduate students and ten engineers. He was interested in the publication preference of scholars in the high-performance computing (HPC) area. He selected a young scholar $S6$ and another scholar $S7$ at his mid-career stage. He first studied $S7$'s highly-cited papers (**T1**). He grouped the papers by "P. Citation" and "Individual Paper" attributes. He hovered the papers in the high-citation group and found that the papers showed diverse topics. He added an extra level of grouping using "P. CCF Rank" between these two levels. He found that only one of his eight high-citation papers was published in a Rank A venue. $E2$ commented that Rank A venues did not guarantee higher numbers of citations. He removed the grouping levels other than "P. CCF Rank" and found that $S7$ received 402 citations from 33 Rank A papers, and 580 citations from 37 Rank B papers, leading to a lower average number of citations for Rank A papers than Rank B papers.

Then, $E2$ added $S6$ to the hierarchical histogram for comparison with $S7$ (**T4**). He found that $S6$, a young Chinese scholar, had a higher tendency to publish in Rank A venues (11 out of 28) than the currently more established scholar $S7$ (33 out of 175). He commented that this might demonstrate the impact of government policies as publishing in Rank A venues was more rewarding in China. He further studied the publication tendency of $S6$ with and without his co-advisors ($S8$ and $S9$) by comparing $S6 + (S8|S9)$ and $S6 - S8 - S9$ (**T3**). He applied alignment to enhance visual comparison. $E2$ found that both paper sets had a similar number of papers of Rank A. However, starting from 2011 (which was only two years after his first paper), $S6$ received more citations from his independent work than from collaborated works with his advisors, as shown in Figure 9 (a).

$E2$ was interested in the publication tendency difference between China and other countries. He added one of the most distinguished scholars in HPC ($S10$ from the United States) for further verification (**T1**). He found that $S10$ only had 42 out of 357 papers published in Rank A venues. Nevertheless, when he studied the tendency over the years, he found that $S10$ did publish more Rank A papers after 2000 (36 out of 235) than before (6 out of 122), as shown in Figure 9 (b).

## 8.3 Discussion and Lessons Learned

Overall, we found that the two experts could form meaningful combinations of scholars and partition the data using appropriate attributes to support their analysis. But the learning curve seemed to vary across these two experts. $E1$ was able to manipulate this graphical tool quickly, even for those relatively complicated functions. For example, it was to our surprise that he used grouping to remove undesired periods instead of brushing in the publication view. $E2$ seemed to experience a steeper learning curve to use the interface. For example, we observed once that he could formulate the combinations of scholars appropriately. It took him some time to find the button to add the paper set into the publication view. This situation improved during the evaluation as he could perform the most desired functions smoothly. The user study with graduate students also indicated that the learning effort was reasonable. After a brief introduction and initial exploration, most students could answer complicated questions using the tool with high accuracy (97%).

However, we should note that while the flexibility supports more tasks, it also leads to confusion of users as they have too many choices. For example, $E1$ missed the opportunity to discover why the year 2009 became a turning point for $S1$ in Figure 8 (d). He suspected that $S1$ was already famous in 2012, which brought more citations to his previous papers. But $S1$'s papers published before 2009 might benefit less from this. However, by replacing the top attribute with "Individual Paper", we found that most papers published by $S1$ in 2009 received fewer citations than $S1$'s 2008 papers. This fact was ignored by $E1$ during the evaluation. $E2$ suggested that we might develop a recommendation system to automatically suggest interesting operations for users or provide templates to combine scholars.

Other limitations of our current implementation were also identified. For example, $E1$ pointed out that our data lacked the papers in the most recent three years. $E2$ was willing to explore the topics of papers authored by $S7$, but $SD^2$ did not incorporate this attribute. He also found the average number of citations interesting in some cases, but $SD^2$ could only demonstrate the numbers of papers or citations. These issues are mostly related to the data set instead of our tool's design, and they have not been addressed yet. Some suggestions are related to the interactions. For example, $E1$ suggested that users should be able to change the operators to combine scholars in the publication view, and $E2$ suggested that users should be able to determine the criteria for high/medium/low citations. We have already incorporated the first feature and will add the second in the future.

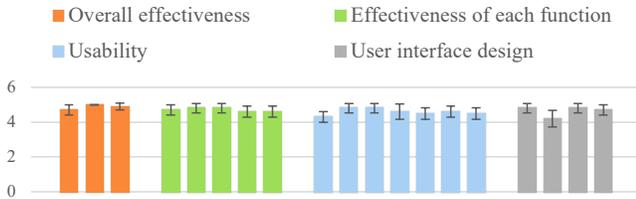

Fig. 10: The rating scores (1-5) received in the first user study. The height of a bar encodes the average score for a corresponding rating question, and the error bar shows the 95% confidence interval.

### 8.4 First Round User Study

After the revision of $SD^2$ based on feedback from the pivotal expert evaluation, we recruited ten graduate students in computer science and performed a user study to evaluate its effectiveness.

**Experimental procedure.** The first user study started with an introduction session where participants were given a briefing of the basic usage and core functions of $SD^2$, including adding scholars, adding paper sets for selection, selecting paper sets for investigation, and operating the hierarchical histogram. After the introduction, participants took a practice session so that they would get familiar with the $SD^2$ interface and interactions. After that, the main study commenced. Participants were asked to perform a list of tasks and answer ten questions. The tasks require them to create paper sets of Jiawei Han and his two students

(Yizhou Sun and Xifeng Yan), Christos Faloutsos, and his two students (Jure Leskovec and Hanghang Tong), for subsequent exploration and investigation using the hierarchical histogram. To answer the questions correctly, participants needed to leverage several essential functions designed for the hierarchical histogram (e.g., changing and reordering grouping attributes, enabling bar alignment, and performing bar grouping). In the end, participants completed a questionnaire. The questionnaire asked them to rate $SD^2$ on a five-point Likert scale (1: *strongly disagree*; 2: *disagree*; 3: *neutral*; 4: *agree*; and 5: *strongly agree*) under four categories: *overall effectiveness*, *individual functions*, *usability*, and *user interface design*. It also asked open questions soliciting detailed comments regarding $SD^2$ and suggestions for improvement. The entire study took approximately one hour for most participants, and each participant was paid $15 as compensation.

**Results and discussion.** The participants achieved very high accuracy (97%) in answering the task questions: seven participants answered all questions correctly, and three participants had only one wrong answer. Considering that most tasks required three to five operations to perform, the accuracy is beyond our expectations. *This demonstrated that $SD^2$ was effective and could be learned in a reasonable amount of time (around half an hour).*

Then, we examined the rating questions. The average score and 95% confidence of intervals are shown in Figure 10. We can see that the participants provided positive ratings for all rating questions. For effectiveness, we found that the overall effectiveness had higher average ratings (ranging from 4.7 to 5 indicated by the orange bars) than that of the individual functions (ranging from 4.6 to 4.8 indicated by the green bars). But for usability, the trend was different. The overall usability received a lower average rating (4.3 indicated by the first light blue bar) than that of the individual functions (ranging from 4.5 to 4.8 indicated by the other blue bars). A possible explanation is that although $SD^2$ can perform the tasks effectively as a whole, none of the individual functions can be used to complete the tasks on their own. Although each of the individual functions was easy to learn, understanding the entire system was more difficult. But given the confidence interval, the difference may not be significant.

For the "user interface design", we find that most scores are close to 5, and the only exception is the statement "the fonts (including color and size) are appropriate to distinguish among different items" (with an average score of 4.2 indicated by the second gray bar). Two participants provided a neutral rating for this statement, and one of them stated in her answer to the open question, "what one change to $SD^2$ would you suggest" that the fonts may be used to distinguish citation attributes and paper attributes in the attribute list.

The comments for the open questions are positive as well. Typical comments includes "*easy to understand and use*", "*flexible*", "*intuitive*", "*user friendly*", and "*runs fast*". One participant mentioned that she "do find the tool requires some learning effort, but it only takes a few minutes." For the question regarding the most useful function, the participants seemed to have different choices. The most popular candidates were the hierarchical histograms (chosen by three participants) and the set operations (chosen by two participants). For the "one change to recommend", most suggestions were related to the interface design, including distinguishing citation/publication attributes, providing more tooltips, and providing a manual. These suggestions were adopted and reflected in a later version of SD², before the pivotal expert evaluation is performed.

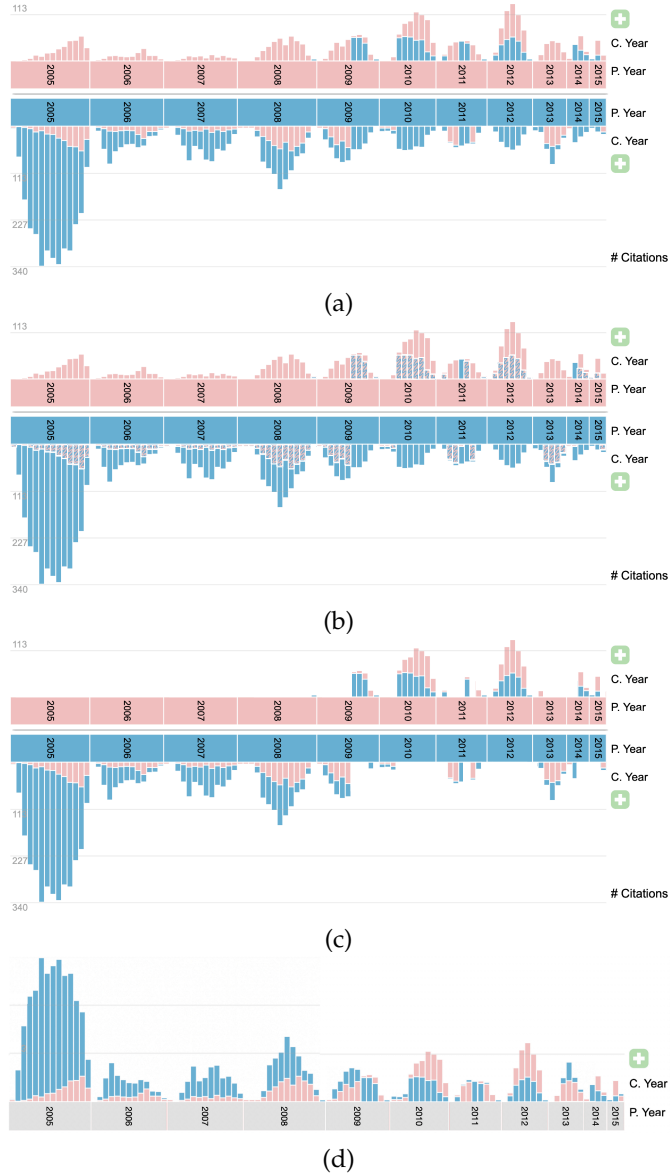

Fig. 11: Design alternatives for the bar alignment examined in the second user study. (a) shows the original two-sided design with solid color filling (TSO+SL). (b) shows the two-sided design with stripped color filling (TSO+SP). (c) shows the two-sided design with shorter bars removed and solid color filling (TSC+SL). (d) shows the one-sided design with solid color filling (OS+SL).

## 8.5 Second Round User Study

The second user study was conducted specifically to improve the design of bar alignment and test the effectiveness of the bar grouping feature. For bar alignment, we created six design alternatives for the participants to identify the one that was most effective and easiest to understand. Two design factors are considered in the alignment: bar placement and color filling. For the bar placement, three options are given: the original two-sided design (TSO), which flips the shorter bar to the other side and overlaying with the aligned longer bar; the two-sided design with cutoff (TSC), which flips the shorter bar to the other side as the TSO does, but removes the shorter bar; the one-sided design (OS), which places all bars on the upper histogram with the shorter ones on the top of the corresponding longer ones. We do not consider the clustered histogram layout in this study, because it may not use the horizontal space efficiently. Please refer to Section 6.3 for a detailed discussion. For the color filling, we offer two options: solid filling (SL), which fills the shorter bar with solid color; and stripped filling (SP), which fills the shorter bar with strips to indicates that the bars are overlaying instead of stacked. Please note that for all alternatives, we enforce a minimum height for each bar, so that it can be read as a very small number.

**Experimental procedure.** The second user study was conducted in two stages. The first stage was a pivotal study that identified the two most promising candidates from the six design choices. Then, a formal study was performed to identify the final design choice from the two candidates based on the same tasks in the first round of user study.

The pivotal study was used to avoid the overwhelming choices in performing the tasks. We recruited thirteen graduate students as participants for the pivotal study. After a brief introduction to the hierarchical histogram, the participants simply read the screenshots of each design choice and rated different factors on a five-point Likert scale in two aspects: clarity (clear to understand) and readability (easy to read). In addition, they were asked to list the three most effective design choices in order. The introduction was given onsite, while the questionnaire was filled online with an anonymous voting system.

For the formal study, we recruited the ten graduate students who participated in the first round user study. They were asked to repeat the tasks in the first study using the two design choices identified in the pivotal study. They were informed that the accuracy and timing would not be recorded. After performing the tasks, they were asked to rate the two choices in terms of clarity and readability. To avoid bias, the statements were provided in two opposite manner: the first stated in the form that "A is better than B" and the second stated in the form that "B is better than A". The scores were then converted in a consistent manner. In addition, four questions were asked to rate the recent changes for the grouping function.

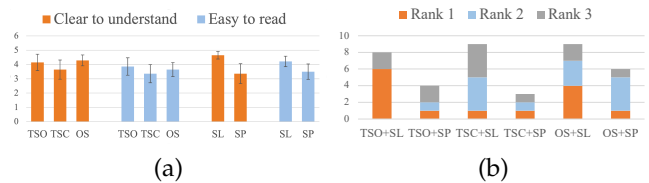

Fig. 12: The results of the pivotal study in the second round user study. (a) shows the rating results, and (b) shows the ranking results.

**Results and discussion.** For the pivotal study, the rat-

 

ing and ranking showed consistent results, as shown in Figure 12. In terms of the color filling design, SL received higher scores than SP with smaller confidence intervals in both clarity and readability. The ranking results showed a similar trend that the SL received higher ranks than the corresponding SP with the same placement design. In terms of the bar placement, TSC received the lowest scores with the largest confidence interval in both clarity and readability. This reflected a more diverse opinion against TSC than the other two placement designs. The ranking results also indicated that TSC was least favored, with only two first-choice votes and twelve votes in the top three. In comparison, TSO received seven votes as the first choice and twelve votes in the top three, and OS received five votes as the first choice and fifteen votes in the top three. Overall, the results showed that TSO and TSC were favored over TSC, and SL was favored over SP. Therefore, we used TSO+SL and OS+SL in the formal study.

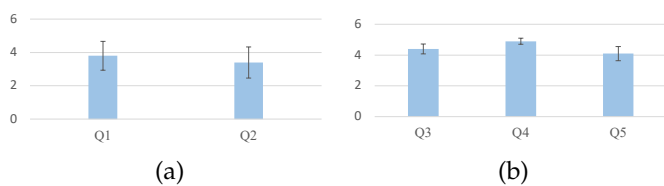

Fig. 13: The results of the formal study in the second round user study. (a) shows the rating scores comparing TSO+SL and OS+SL. (b) shows the rating scores for the updated grouping feature.

In the formal study, OS was slightly favored over TSO in both clarity and readability, although the difference was marginal, as shown in Figure 13 (a). For clarity, OS was considered to outperform TSO with an average score of 3.8 and a relatively large confidence interval. For readability, the difference is even smaller with an average score of 3.4. The participants supporting OS stated that it allowed the numbers to be read on one side, which was easier to track the trends of both the upper and lower histograms. Besides, they mentioned that TSO might be confusing as the red and blue bars appeared on both sides. On the other hand, the main reason for the participants supporting TSO was that it was more consistent with the unaligned view. Investigating into the detailed scores, we found that while eight scores strongly supported OS (with a score of 5), only one strongly supported TSO (with a score of 1). In addition, twelve scores favored OS (with a score of 4 or 5), while only seven scores favor TSO (with a score of 1 or 2). Therefore, although participants with both the two different opinions provided reasonable arguments, we used OS+SL in our final version of $SD^2$ based on the rating scores.

**For the grouping function,** most participants stated that it was easy to understand (4.4) and convenient to use (4.9). All participants stated in an open question that they did use the grouping function performing the tasks. The most common task for them to apply the grouping is the comparison of the scholarly outcomes in two periods of time. One participant also mentioned that she used the grouping function to reduce the visual complexity. However, for the newly added "ignore" button, the rating was slightly lower (4.1). The participants stated that the tasks might not necessarily require a group to be ignored. Instead, as the grouping already provided a relatively concise visualization, they could often simply focused on the other groups without explicitly removing a specific one.

## 9 COMPARISON WITH EXISTING TECHNIQUES

We compare our $SD^2$ with several existing techniques based on the core requirements in Section 4: **R2**, **R3**, and **R4**. We do not consider **R0** and **R1** in this comparison, as these two requirements specify that visual hints should be provided to select attributes and scholars, which are less relevant to the core tasks. We compare $SD^2$ with three recent egocentric techniques (i.e., ImpactVis [32], Influence Flower [26], and VisPubComPAS [33]), a cross-filtering design (i.e., Keshif [37]), and a popular website (i.e., AMiner [1]).

**For the combination of scholars**, we compare these techniques based on their supported combination operations and the number of scholars to combine. These two factors largely decide what kind of relationships among scholars can be analyzed. ImpactVis and AMiner show the number of collaborated papers of a scholar and each of her top coauthors. Influence Flower allows an institution to be studied as an entity. But it does not allow users to freely specify scholars to form a group as VisPubComPAS does. The techniques mentioned above only provide a single way to combine scholars. Therefore, they only allow a specific relationship (e.g., coauthorships) to be discovered. Both Keshif and $SD^2$ support three set operations (i.e., "and", "or", and "not") to flexibly combine the scholars and derive various kinds of relationships.

**For information partition and organization**, we compare these techniques based on the number of maximum partition levels and whether the attributes used for partition can be customized. These two factors determine the interactions among which attributes can be discovered. Most of these techniques except Influence Flower and AMiner support information partition. ImpactVis uses a three-level partition using three fixed attributes (publication and citation year, paper types). But for the coauthors, ImpactVis only partitions the collaborated papers by their topics. VisPubComPAS provides a one-level partition using a single fixed attribute. Keshif allows the data to be partitioned using a user-specified attribute (e.g., venues and authors). Keshif further employs cross-filtering to investigate the relationships among multiple attributes. But we should note that cross-filtering does not provide a full picture of the interactions among all attributes immediately. Users need to interact with the histograms to discover details. $SD^2$ provides the maximum partitioning flexibility among these techniques, as it allows users to freely select any number of attributes for partition.

**For comparison**, most of these techniques, except ImpactVis and AMiner, allows the data from two (groups of) scholars to be visually compared. Influence Flower uses a node-link diagram representation and displays an overlaid background diagram for comparison. VisPubComPAS, Keshif, and our $SD^2$ are based on bar charts, and they align the corresponding bars for comparison. More specifically, VisPubComPAS and Keshif use clustered bars which is scalable to more than two entities. Our $SD^2$ aligns upper-lower hierarchical histograms, which cannot support comparisons of more than two entities.

Overall, our SD$^2$ is the most appropriate one to support our tasks as it fulfills all requirements. Other techniques might not be able to perform some tasks when they fail to meet certain requirements.

## 10 CONCLUSIONS AND FUTURE WORK

We have presented SD$^2$, a visual analytics tool for visualizing and comparing scholarly performance using papers and citations. SD$^2$ addresses the need for objectively and comprehensively evaluating the scholarly output and impact of researchers. It allows users to understand a researcher's scholarly performance via publication and citation records, evaluate the independence of a researcher, compare individual researchers or groups of researchers. Following the interaction model of "*search, show context, expand on demand*", SD$^2$ provides a suite of user interactions that target real-world scenarios where the evaluators or recruiters know who to search or compare. We demonstrate the effectiveness of SD$^2$ with several case studies that are driven by common needs of scholarly performance evaluation. SD$^2$ has also been assessed via expert evaluations, followed by a user study. In the future, we will seek to expand the data used by SD$^2$ from CS-related fields to other science and engineering fields.

**Generalization.** The exploration scheme and the hierarchical histogram introduced in SD$^2$ are not limited to visualizing scholarly data. They can be generalized and applied to other faceted, categorical data. Toward this end, the core tasks can be abstracted as: **set combination, information partitioning**, and **comparison**. The set combination task allows users to combine multiple sets of elements (e.g., belonging to different categories or related to different subjects, etc.) in their desired manner. The information partitioning task slices and dices the data according to their attributes and visualizes the results in a hierarchical histogram. The comparison task allows users to visually compare two hierarchical histograms. This may help to answer questions that are not limited to science and engineering fields. For example, when archaeologists study cultural relics discovered in different ruins, they may be interested in: *what are the relics found at this site but not at that one* (set combination), *how many relics are found of each type, how the size of relics distributes of each type* (information partitioning), and *how the relics different from two sites* (comparison)? Our workflow may help to answer these questions and facilitate the understanding of different cultures. Similarly, the exploration scheme can be applied to other types of data where labels are available to group elements into sets. For example, the posts on social media are often associated with attributes such as users, time, locations, lengths, and tags. We can use the same scheme to build hierarchical sets from the social media data and explore the interactions among different attributes by visually comparing the sets.

## ACKNOWLEDGEMENTS

This research was supported in part by the National Natural Science Foundation of China through grant 61902446 and 91937302, Key R&D Program of Guangdong through grant 2019B010151001, and the U.S. National Science Foundation through grants IIS-1455886, DUE-1833129, IIS-1955395, IIS-2101696, and OAC-2104158. The authors would like to thank Sichen Jin who contributed to the initial development of this work.

## REFERENCES

[1] AMiner. AI-10 most influential scholar lists. https://www.aminer.cn/ai10/, 2018.

[2] F. Beck, F.-J. Wiszniewsky, M. Burch, S. Diehl, and D. Weiskopf. Asymmetric visual hierarchy comparison with nested icicle plots. In *Joint Proceedings of International Workshop on Euler Diagrams and Graph Visualization in Practice*, pages 53–62, 2014.

[3] N. Cao, Y. R. Lin, F. Du, and D. Wang. Episogram: Visual summarization of egocentric social interactions. *IEEE Computer Graphics and Applications*, 36(5):72–81, 2016.

[4] China Computer Federation. A list of computer science conferences and journals recommended by CCF. http://www.ccf.org.cn/xspj/gym1/, 2019.

[5] M. Dork, N. H. Riche, G. Ramos, and S. Dumais. PivotPaths: Strolling through faceted information spaces. *ACM Transactions on Interactive Intelligent Systems*, 18(12):2709–2718, 2012.

[6] F. Du, C. Plaisant, N. Spring, and B. Shneiderman. Visual interfaces for recommendation systems: Finding similar and dissimilar peers. *ACM Transactions on Intelligent Systems and Technology*, 10(1):9:1–9:23, 2019.

[7] T.-L. Fung, J.-K. Chou, and K.-L. Ma. A design study of personal bibliographic data visualization. In *Proceedings of IEEE Pacific Visualization Symposium*, pages 244–248, 2016.

[8] M. Glueck, P. Hamilton, F. Chevalier, S. Breslav, A. Khan, D. Wigdor, and M. Brudno. PhenoBlocks: Phenotype comparison visualizations. *IEEE Transactions on Visualization and Computer Graphics*, 22(1):101–110, 2015.

[9] J. A. Guerra-Gómez, A. Buck-Coleman, M. L. Pack, C. Plaisant, and B. Shneiderman. TreeVersity: Interactive visualizations for comparing hierarchical data sets. *Transportation Research Record*, 2392(1):48–58, 2013.

[10] F. Heimerl, Q. Han, S. Koch, and T. Ertl. CiteRivers: Visual analytics of citation patterns. *IEEE Transactions on Visualization and Computer Graphics*, 22(1):190–199, 2016.

[11] P. Isenberg, F. Heimerl, S. Koch, T. Isenberg, P. Xu, C. Stolper, M. Sedlmair, J. Chen, T. Möller, and J. Stasko. vispubdata.org: A metadata collection about IEEE visualization (VIS) publications. *IEEE Transactions on Visualization and Computer Graphics*, 23(9):2199–2206, 2017.

[12] J. B. Kruskal and J. M. Landwehr. Icicle plot: Better displays for hierarchical clustering. *The American Statistician*, 37(2):162–168, 1983.

[13] J. Lamping and R. Rao. The hyperbolic browser: A focus+context technique for visualizing large hierarchies. *Journal of Visual Languages and Computing*, 71(1):33–55, 1996.

[14] B. Lee, M. Czerwinski, G. Robertson, and B. B. Bederson. Understanding research trends in conferences using PaperLens. In *Proceedings of ACM CHI Extended Abstracts on Human Factors in Computing Systems*, pages 1969–1972, 2005.

[15] B. Lee, G. G. Robertson, M. Czerwinski, and C. S. Parr. CandidTree: Visualizing structural uncertainty in similar hierarchies. *Information Visualization*, 6(3):233–246, 2007.

[16] V. I. Levenshtein. Binary codes capable of correcting deletions insertions and reversals. *Soviet Physics—Doklady*, 10(8):707–710, 1966.

[17] A. Lex, N. Gehlenborg, H. Strobelt, R. Vuillemot, and H. Pfister. UpSet: Visualization of intersecting sets. *IEEE Transactions on Visualization and Computer Graphics*, 20(12):1983–1992, 2014.

[18] G. Li, Y. Zhang, Y. Dong, J. Liang, J. Zhang, J. Wang, M. J. McGuffin, and X. Yuan. BarcodeTree: Scalable comparison of multiple hierarchies. *IEEE Transactions on Visualization and Computer Graphics*, 26(1):1022–1032, 2019.

[19] D. Liu, F. Guo, B. Deng, H. Qu, and Y. Wu. egoComp: A node-link-based technique for visual comparison of ego-networks. *Information Visualization*, 16(3):179–189, 2017.

[20] J. Liu, T. Tang, W. Wang, B. Xu, X. Kong, and F. Xia. A survey of scholarly data visualization. *IEEE Access*, 6:19205–19221, 2018.

[21] Z. Liu, S. H. Zhan, and T. Munzner. Aggregated dendrograms for visual comparison between many phylogenetic trees. *IEEE Transactions on Visualization and Computer Graphics*, 26(9):2732–2747, 2020.

[22] E. Maguire, J. M. Montull, and G. Louppe. Visualization of publication impact. In *Proceedings of Eurographics / IEEE VGTC Conference on Visualization: Short Papers*, 2016.

[23] T. Munzner, F. Guimbretière, S. Tasiran, L. Zhang, and Y. Zhou. TreeJuxtaposer: Scalable tree comparison using focus+context with guaranteed visibility. *ACM Transactions on Graphics*, 22(3):453–462, 2003.

[24] A. Rind, A. Haberson, K. Blumenstein, C. Niederer, M. Wagner, and W. Aigner. PubViz: Lightweight interactive visualization of publication data. In *Proceedings of Eurographics / IEEE VGTC Conference on Visualization: Short Papers*, 2017.

[25] L. Shi, H. Tong, J. Tang, and C. Lin. VEGAS: Visual influEnce GrAph Summarization on citation networks. *IEEE Transactions on Knowledge and Data Engineering*, 27(12):3417–3431, 2015.

[26] M. Shin, A. Soen, B. T. Readshaw, S. M. Blackburn, M. Whitelaw, and L. Xie. Influence flowers of academic entities. In *Proceedings of IEEE Conference on Visual Analytics Science and Technology*, pages 1–10, 2019.

[27] B. Shneiderman. Tree visualization with tree-maps: A 2D space-filling approach. *ACM Transactions on Graphics*, 11(1):92–99, 1992.

[28] J. Stasko and E. Zhang. Focus+context display and navigation techniques for enhancing radial, space-filling hierarchy visualization. In *Proceedings of IEEE Symposium on Information Visualization*, pages 57–65, 2000.

[29] Y. Tu and H.-W. Shen. Visualizing changes of hierarchical data using treemaps. *IEEE Transactions on Visualization and Computer Graphics*, 13(6):1286–1293, 2007.

[30] F. van Ham and A. Perer. Search, show context, expand on demand: Supporting large graph exploration with degree-of-interest. *IEEE Transactions on Visualization and Computer Graphics*, 15(6):953–960, 2009.

[31] R. van Liere and W. de Leeuw. GraphSplatting: Visualizing graphs as continuous fields. *IEEE Transactions on Visualization and Computer Graphics*, 9(2):206–212, 2003.

[32] Y. Wang, C. Shi, L. Li, H. Tong, and H. Qu. Visualizing research impact through citation data. *ACM Transactions on Interactive Intelligent Systems*, 8(1):5:1–5:24, 2018.

[33] Y. Wang, M. Yu, G. Shan, H.-W. Shen, and Z. Lu. VisPubComPAS: A comparative analytical system for visualization publication data. *Journal of Visualization*, 22(5):941–953, 2019.

[34] W. Willett, J.Heer, and M. Agrawala. Scented widgets: Improving navigation cues with embedded visualizations. *IEEE Transactions on Visualization and Computer Graphics*, 13(6):1129–1136, 2007.

[35] M. Q. Y. Wu, R. Faris, and K.-L. Ma. Visual exploration of academic career paths. In *Proceedings of IEEE/ACM International Conference on Advances in Social Networks Analysis and Mining*, pages 779–786, 2013.

[36] Y. Wu, N. Pitipornvivat, J. Zhao, S. Yang, G. Huang, and H. Qu. egoSlider: Visual analysis of egocentric network evolution. *IEEE Transactions on Visualization and Computer Graphics*, 22(1):260–269, 2016.

[37] M. A. Yalcin, N. Elmqvist, and B. B. Bederson. Keshif: Rapid and expressive tabular data exploration for novices. *IEEE Transactions on Visualization and Computer Graphics*, 24(8):2339–2352, 2018.

[38] J. Zhang, C. Chen, and J. Li. Visualizing the intellectual structure with paper-reference matrices. *IEEE Transactions on Visualization and Computer Graphics*, 15(6):1153–1160, 2009.

[39] J. Zhao, M. Glueck, F. Chevalier, Y. Wu, and A. Khan. Egocentric analysis of dynamic networks with EgoLines. In *Proceedings of ACM SIGCHI Conference*, pages 5003–5014, 2016.

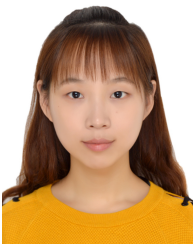
**Zhichun Guo** is a Ph.D. student at the University of Notre Dame. She received a BS degree in computer science from Fudan University in 2019. Her research interests are graph representation learning and visual analytics.

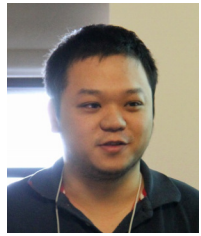
**Jun Tao** is an associate professor of computer science at Sun Yat-sen University and National Supercomputer Center in Guangzhou. He received a Ph.D. degree in computer science from Michigan Technological University in 2015. His major research interest is scientific visualization, especially on applying information theory, deep learning, and optimization techniques to interactive flow visualization and multivariate data exploration.

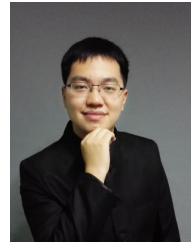
**Siming Chen** is an associate professor at Fudan University. Prior to this, he was a research scientist at Fraunhofer Institute IAIS and a postdoctoral researcher at the University of Bonn in Germany. He received a Ph.D. degree from Peking University. His research interests include visual analytics of social media, cybersecurity, and spatiotemporal data. He has published several papers in IEEE VIS, IEEE TVCG, EuroVis, etc. More information can be found at http://simingchen.me.

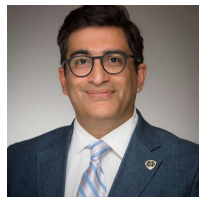
**Nitesh V. Chawla** is the Frank M. Freimann Professor of Computer Science and Engineering at the University of Notre Dame. He received a Ph.D. degree in computer science and engineering from the University of South Florida in 2002. He is the Founding Director of the Lucy Family Institute for Data and Society. His research is focused on machine learning, data science, and network science, and is motivated by the question of how technology can advance the common good through interdisciplinary research.

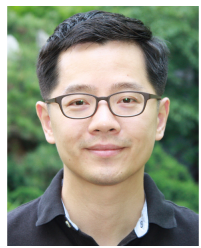
**Chaoli Wang** is an associate professor of computer science and engineering at the University of Notre Dame. He received a Ph.D. degree in computer and information science from The Ohio State University in 2006. Dr. Wang's main research interest is data visualization, in particular on the topics of time-varying multivariate data visualization, flow visualization, as well as information-theoretic algorithms, graph-based techniques, and deep learning solutions for big data analytics. He is an associate editor of IEEE Transactions on Visualization and Computer Graphics.

# APPENDIX

## 1 FEEDBACK FROM PIVOTAL EXPERT EVALUATION

We evaluated the prototype of SD² with four experts. Two experts ($E3$ and $E4$) are postdoctoral researchers specializing in visual analytics. Both of them have more than five years of experience and publish multiple papers at top venues, such as TVCG, VAST, and CHI. Expert $E5$ is a senior applied scientist at Microsoft Research. His research focuses on social and information networks and data mining, with years of experience in analyzing MAG, which is a major data source of SD². Expert $E6$ is a full professor of computer science specializing in bioinformatics, whose department has over a hundred faculty members. He is experienced in recruiting faculty members and postdoctoral researchers. The pivotal expert evaluation was performed in the form of interviews, including a brief introductory section and a free exploration section. The interview took around one hour each, and the experts were asked to provide textual feedback after that. Based on their feedback, we revised our tool accordingly. The feedback and revision procedure repeats in multiple rounds, and the details are reported in Section 2 of the Appendix.

**Overall sentiment.** The feedback received is generally favorable. The experts confirm that SD² is intuitive and useful. $E6$ commented that "*SD² shows different aspects of scholar data clearly, and it is easy to interpret.*" and $E4$ stated that "*I believe it is useful for evaluating the scholars' publication records in depth. From the recruiting perspective, it would be especially useful.*" The experts were particularly interested in our functions that combine scholars for comparative investigation. $E3$ commented that "*I really like the set-operation based filtering, which is quite intuitive and useful.*" and $E4$ said that "*The powerful function of the tool is the flexible logical combination for customized filtering and comparing. Normal tools don't support that. And viewing the records from publication/citation perspectives also bring new insights.*"

**Key comments.** The experts also made the following comments in terms of tool functionality, scalability, and generalization. For comparing the performance of two scholars, $E3$ commented that "*the comparison shows exciting results. The bar alignment is especially useful, which makes the difference much easier to perceive.*" For interactive filtering functions, $E4$ commented that "*If possible, period selection for citation years can be added, it can enable one to drill down to specific years.*" and "*If individual records' citation details can also be shown, it would be perfect.*" We can handle the filtering through grouping in our current implementation. However, we admit that it is not convenient, and a better strategy should be sought. For using the tool to analyze large-scale effects or general trends in a field, $E5$ commented that "*I would like to use this tool to explore recent trends in my area. For example, using this tool to study whether the researchers in the data mining fields are citing more machine learning papers than database papers nowadays.*" $E6$ mentioned the desire to "*analyzing a larger scope of the academic network by automatically including more related scholars and papers.*" These comments show the willingness of the experts in using SD² for various purposes, which is quite encouraging. Further development effort is needed to enable the tool to handle larger graph structures.

For tool deployment and generalization, $E5$ commented that "*It would be very useful to make the tool publicly available.*" $E6$ (the expert in bioinformatics) added that "*It might be generalized to deal with the relationships between proteins and genes in biomedical engineering as well.*" These comments also indicate that the experts are very interested in the tool.

Finally, we excerpt additional comments from the experts as follows. $E3$: "*I think this tool would be very helpful for many people, including students who want to find a suitable advisor, researchers who want to find suitable collaborators, as well as university administrators who want to recruit new faculties and evaluating the performance of the current faculties, etc.*" "*It takes some time to learn how to use the tool, especially the detailed interactions of those buttons [in the histogram view], but overall it is easy to use and quite flexible.*" "*The case study of Julian McAuley is quite interesting. It definitely shows the usefulness of the tool. Jure is famous in the field of graph mining and graph neural networks, etc. This case may provide support for the empirical observation that a well-established/famous co-author can boost the paper citation, probably due to their good reputation.*" $E4$: "*Individual researchers will be interested in his/her own records. I believe it can provide some new insights. With the tool, people can find their 'majority audience' and communities; even maybe some they didn't notice. For example, in this case, Jian Zhao may [want to] be involved more in the IUI conference, if he didn't do it before.*" $E5$: "*The way to combine multiple scholars is inspiring.*" "*The interaction is very smooth, especially when all data are loaded.*" $E6$: "*The tool performs the proposed tasks well, especially in terms of evaluating the independence of a researcher. This can be very helpful when recruiting new faculty members.*" "*What I find more interesting is the attributes of citation data. This assists in the evaluation of citation quality, which is critical to evaluate one's research quality. It is important to know whether a young researcher is recognized by renowned experts in high-quality papers.*" "*I feel the tool is flexible to be extended in more scenarios. For example, analyzing a larger scope of the academic network by automatically including more related scholars and papers.*"

## 2 TOOL REVISION

We revise our tools in multiple passes by addressing the issues identified from the pivotal expert evaluation and user study, and the ones raised from the previous review cycle. The interview was conducted two or three times for each expert to confirm that their concerns are addressed.

First, the original version of SD² provided a simple slider for users to pan the hierarchical histogram back and forth along the horizontal direction. The revision implements a *mini-map* feature (refer to Figure 1). With the mini-map, users can specify a range and drag the resulting time window back and forth for more convenient temporal exploration of the hierarchical histogram, while keeping the entire period insight.

Second, besides the standard *linear* scale provided for displaying the vertical bars showing the corresponding quantities in the hierarchical histogram, we add the *square root* and *logarithmic* scales for users to choose. These two new scales are useful for visual comparison of different bars showing a mix of rather high and rather low quantities.

Third, initially, we used the same gray color for showing the vertical bars in both the upper and lower histograms.

With such a uniform color scheme, experts raised a concern that it is not easy to observe bar alignments. Therefore, we switch to a pink vs. blue color scheme that shows the upper and lower histograms for easy alignment and clear contrast. In addition, we further overlay the shorter bar on top of the corresponding taller bar, so that their difference can be easily evaluated visually (refer to, for example, Figure 3 (b)).

Fourth, our earlier version does not provide a visual correspondence between the paper sets in the publication view and the hierarchical histograms. A user needs to use the text description to identify the correspondence. An expert pointed out that this may lead to extra effort for users to identify paper sets for investigation. We address this issue by color coding: the paper sets used in the hierarchical histogram view are colored according to the histogram color (i.e., pink or blue), and the other paper sets are colored in green.

Fifth, we also make some minor revisions according to the feedback received from the participants of the user study. These revisions include adding a manual for the interface (which can be accessed by clicking the "help?" button at the top-right corner), adding tooltips to provide operational hints, adding a separation line between the publication and citation attributes, and polishing the style of the buttons.

Sixth, the alternative design choices for bar alignment were studied with an additional round of user study. The study suggested that the original design which duplicated the shorter bars on both sides might not be optimal. After the study, we made a revision to align all bars on the upper side, which received the most positive rating from the participants.

We point out that there are helpful comments that have not been fully implemented yet. First, the expert from the MAG project suggested we use the Azure Academic Knowledge (AAK) API to avoid the issues caused by the anti-crawling mechanism of GS. However, as scholars maintain the publication data at GS, we still consider it to be more reliable for small-scale analysis (such as the case studies in this paper). For the tool to be publicly useful, we believe AAK API will be a more practical solution. Second, an anonymous reviewer suggested that using the tooltip to select an attribute might be prone to misoperations (such as accidentally moving the mouse out of the tooltip). As we do not find an appropriate alternative that is more robust and still maintains the compactness of the interface, we enlarge the detection area of the tooltip, so that it will only disappear when the mouse is a certain distance away.